\documentclass[twocolumn,aps,prl,showpacs]{revtex4}
\usepackage{amsmath}
\usepackage{amssymb}
\usepackage{color}
\usepackage{graphicx}

\begin{document}

\title{Photons with tunable spectral shapes: The transition from frequency
anticorrelated to correlated photon pairs}

\author{M. Hendrych$^1$, M. Mi\v{c}uda$^{1,3}$, A. Valencia$^1$ and J. P.
Torres$^{1,2}$ }

\affiliation{$^{1}$ICFO-Institut de Ci\`{e}ncies Fot\`{o}niques,
Mediterranean Technology Park, Castelldefels (Barcelona),
Spain, \\
$^{2}$Universitat Polit\`{e}cnica de Catalunya, Department of
Signal Theory and Communications, Barcelona, Spain,\\
$^{3}$Department of Optics, Palack\'{y} University, Olomouc, Czech
Republic.}


\begin{abstract}
We present an experimental demonstration of the full control of
the frequency correlations of entangled photon pairs. The joint
spectrum of photon pairs is continuously varied from photons that
exhibit anticorrelation in frequency to photons that exhibit
correlation in frequency, passing through the case of uncorrelated
photons. Highly entangled frequency-anticorrelated photon pairs
were obtained even when an ultrafast laser was used as a pump. The
different kinds of correlations are obtained without changing
neither the wavelength, nor the nonlinear crystal.
\end{abstract}

 \maketitle

\begin{figure}[htb]
\centerline{\scalebox{0.25}{\includegraphics{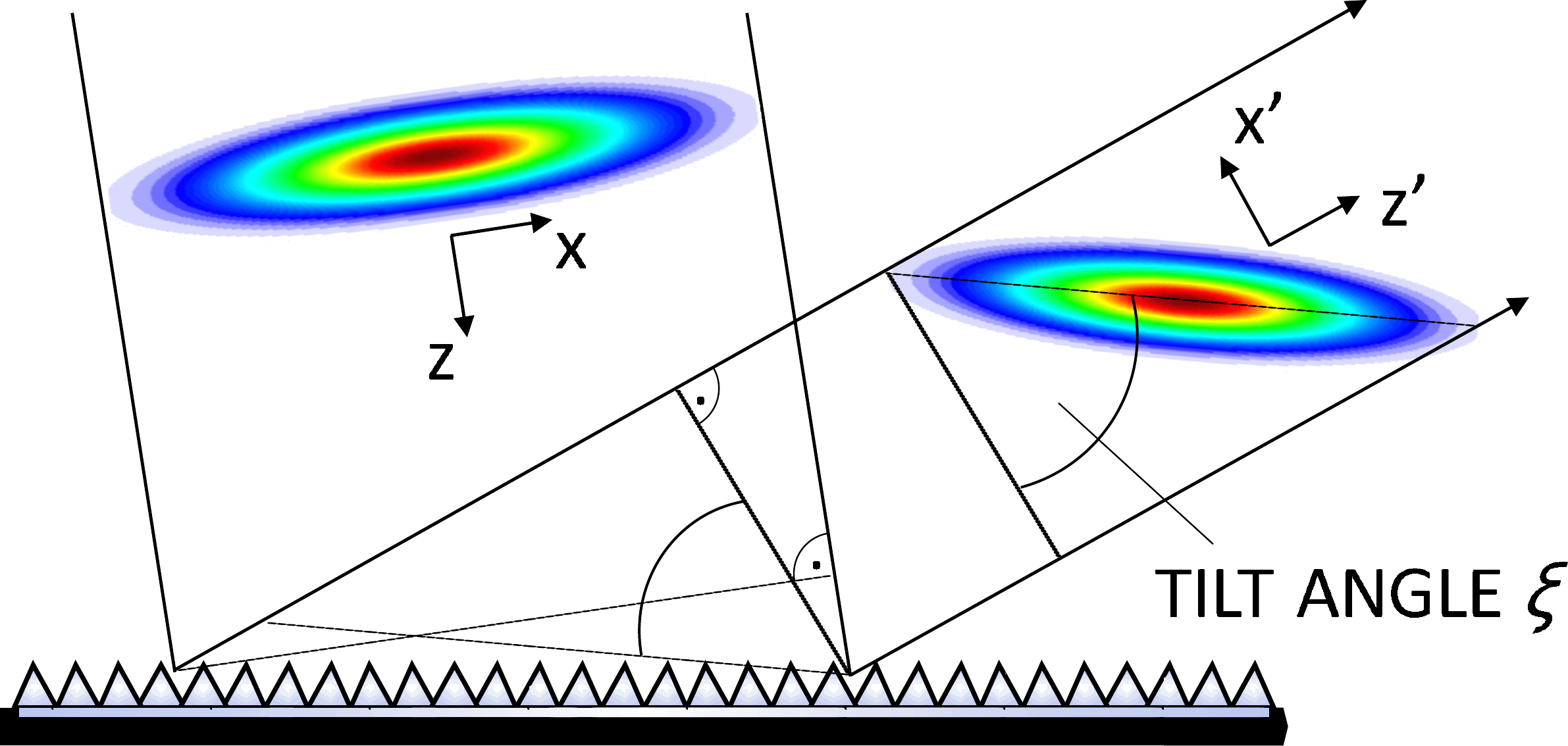}}}
\caption{(color online). Angular dispersion tilts the front of a
pulse by an angle $\xi$. After the element that introduces angular
dispersion (prism or grating), the pulse front is no longer
perpendicular to the direction of propagation. }\label{tilt}
\end{figure}

The motivations to control the joint spectrum of paired photons
come from both, the applied and fundamental points of view.
Frequency-uncorrelated photons are desired because they provide a
source of heralded pure single photons, an important tool in
quantum computing and communications~\cite{singlephotons}. On the
other hand, frequency-correlated and anticorrelated photons are
important for quantum metrology, because these types of
correlations can enhance the precision of some existing time
measurement and synchronization techniques~\cite{metrology}.
Moreover, the possibility to generate symmetric spectral shapes of
entangled photons guarantees that all the distinguishing
information that may come from the spectra of the photons is
erased. This is a key issue when considering applications based on
interferometric techniques, where any distinguishability reduces
the visibility of the interference
pattern~\cite{distinguishability}.
\\
\indent The search for a way to control the joint spectrum of
paired photons has been of paramount importance in the past few
years. In particular, techniques based on the choice of specific
materials and wavelengths have been used to obtain
frequency-uncorrelated~\cite{mosley1} and frequency-correlated
photons~\cite{kuzucu1}. Experiments that allow a tunable control
of the joint spectrum have also been reported~\cite{Martin_OL07,
Ale_PRL07, Walmsley_PRL09}. However, a unique experiment showing
the transition from anti-correlated to correlated frequency
photons passing through the uncorrelated case has not been
reported so far.
\\
\indent In this Letter, the technique reported in
Ref.~\cite{torres1} is used to directly manifest the full control
over the joint spectrum of paired photons generated in spontaneous
parametric down conversion (SPDC). It is possible to generate any
desired type of frequency correlations with the same nonlinear
material and at the same wavelength of the pump laser.
Experimental data that clearly demonstrate the tunability from
frequency-anticorrelated photons to frequency-correlated photons
passing through the case of uncorrelated photons pairs are
reported. The measurement of the joint spectrum is performed
directly by measuring the wavelength of each of the photons that
constitute the pair and therefore a visual and immediate
understanding of the type of frequency correlations can be clearly
seen. Interestingly, the joint spectrum of frequency-correlated
photons and the joint spectrum of frequency-anticorrelated photons
generated with a femtosecond pump are reported. To the best of our
knowledge, it is the first time that the joint spectra for these
two cases are measured directly. The other papers that report the
generation of frequency-correlated photons do not measure the
joint spectrum directly, but rather indirectly deduce the type of
frequency correlations by means of interferometric
techniques~\cite{kuzucu1, Martin_OL07} or by using
frequency-conversion processes to measure in the temporal
domain~\cite{kuzucu2}. It is worth stressing that the correlated
and ultrashort-pulsed frequency-anticorrelated cases are just two
particular possibilities of all the correlations and bandwidths
that can be achieved with the experimental technique reported
here.
\\
\indent Generally, when we are interested in the frequency
properties of paired photons, the light beams are projected into
single modes by, for example, coupling the light into singlemode
fibers. All the information concerning the frequency correlations
of paired photons with frequencies $\omega_{j}$ can thus be
obtained from the joint spectrum, or biphoton, of the two-photon
state, $\Phi(\omega_{1},\omega_{2})$ \cite{distinguishability}.
The joint spectral intensity, which corresponds to the probability
of detecting a photon with frequency $\omega_{1}$ in coincidence
with the other photon with frequency $\omega_{2}$, writes
$S(\omega_{1},\omega_{2})=|\Phi(\omega_{1},\omega_{2})|^2$.
\\
\indent One of the most convenient ways of generating paired
photons is the process of spontaneous parametric down-conversion
in which the interaction of a strong pump beam with a nonlinear
medium occasionally results in the production of two photons known
as the signal and idler, whose energies sum up to that of the pump
photon. When the pump beam is chosen to be a continuous wave beam,
energy conservation imposes frequency anti-correlation between the
downconverted photons. This is not so in the general case when the
nonlinear medium is illuminated by a broadband pump. In this case,
the nature of the frequency correlations of the paired photons is
determined by both, the energy conservation and phase-matching
conditions \cite{distinguishability}. It is precisely the proper
control of the phase-matching conditions what allows us to tune
the type of frequency correlations at will.
\\
\indent The experimental results reported in this Letter measure
the joint spectral intensity, $S \left( \omega_s,\omega_i\right)=
|\Phi \left(\omega_s,\omega_i\right)|^2$, of signal and idler
photons with frequencies $\omega_s$ and $\omega_i$. The two-photon
state at the center of the nonlinear crystal ($z=0$) writes
$|\Psi>=\int d\omega_s d\omega_i \Phi
\left(\omega_s,\omega_i\right) a_s^{\dagger}(\omega_s)
a_i^{\dagger}(\omega_i) |0>_s|0>_i$, where
\begin{equation}
\label{jointspectrum}
    \Phi \left(\omega_s,\omega_i\right)=
    E_p \left(\omega_s+\omega_i\right) \rm{sinc}
    \left(\Delta_k L/2 \right),
\end{equation}
$E_p \left(\omega_s+\omega_i\right)$ is the frequency amplitude
distribution of the pump beam at $z=0$, $L$ is the length of the
crystal, and $\Delta_k = k_p-k_s-k_i$ is the phase-mismatch
function, with $k_j$ being the longitudinal component of the pump,
signal and idler wavevectors, respectively.
\\
\indent The control over the joint spectrum experimentally
reported in this Letter is based on the possibility of modifying
$\Delta_k$ by introducing angular dispersion into the pump and the
downconverted beams. The introduction of angular dispersion
enables us to tune the frequency characteristics of paired photons
by manipulating the transverse momentum. In general, when angular
dispersion is applied to a pulse, it causes the front of the pulse
to be tilted by an angle $\xi$ (see Fig.~\ref{tilt}). It means
that the front of the pulse acquires a temporal delay that depends
on its transversal coordinate~\cite{UltrafastBook}.

The introduction of angular dispersion into the pump and the down
converted photons results in an effective inverse group velocity
$N_j^{\prime}$ and inverse group velocity dispersion
$D_j^{\prime}$ ~\cite{torres2}
\begin{eqnarray}
\label{effectivegv}
    N_j^{\prime} &=& N_j + \tan\xi\,\tan \rho_j/c, \nonumber \\
    D_j^{\prime} &=& D_j + (\tan\xi/c)^2/k_j,
\end{eqnarray}
where $N_j$ and $D_j$ are the inverse group velocity and group
velocity dispersion due to the the material dispersive properties,
$\rho$ denotes the Poynting-vector walk-off angle, and $c$ is the
speed of light. By performing a second-order Taylor expansion,
$\Delta_k$ can be written as a function of the effective inverse
group velocities and group velocity dispersion of the interacting
waves
\begin{eqnarray}
\label{Deltak} \Delta k &=&  (N_{p}'-N_{s}')\Omega_{s} +
(N_{p}'-N_{i}')\Omega_{i} \\
&+& 1/2 (D_p' - D_s')\Omega_{s}^2 + 1/2 (D_p' - D_i')\Omega_{i}^2
+ D_p'\Omega_{s}\Omega_{i}, \nonumber
\end{eqnarray}
where $\Omega_{j}$ is the frequency detuning of signal and idler
photons from the central frequency. The possibility to tune
$\Delta_k$, and thereby the frequency correlations, by introducing
angular dispersion can be understood by inspecting the effective
group velocities $N_j^{\prime}$ and group velocity dispersion
$D_j'$ of Eq.~(\ref{effectivegv}).

The effective dispersive properties of $\Delta k$ arise from two
sources: from the material dispersion of the nonlinear crystal and
from the presence of pulse-front tilt $\xi$. A quantitative
comparison shows that both contributions are of the same order.
For example, at a pump wavelength of $400$ nm, the difference of
group velocities $N_p-N_i$ is equal to $\sim 77$ fs/mm.
Comparatively, a typical value of walk-off $\rho_p \sim 4^{o}$
results in a contribution of the tilt $\xi$ to the effective group
velocity mismatch of $\sim 240 \cdot \tan \xi$ fs/mm. This implies
that the spectral shape of the entangled photons can be greatly
tuned by modifying the amount of angular dispersion.

\begin{figure}[t]
\centerline{\scalebox{0.65}{\includegraphics{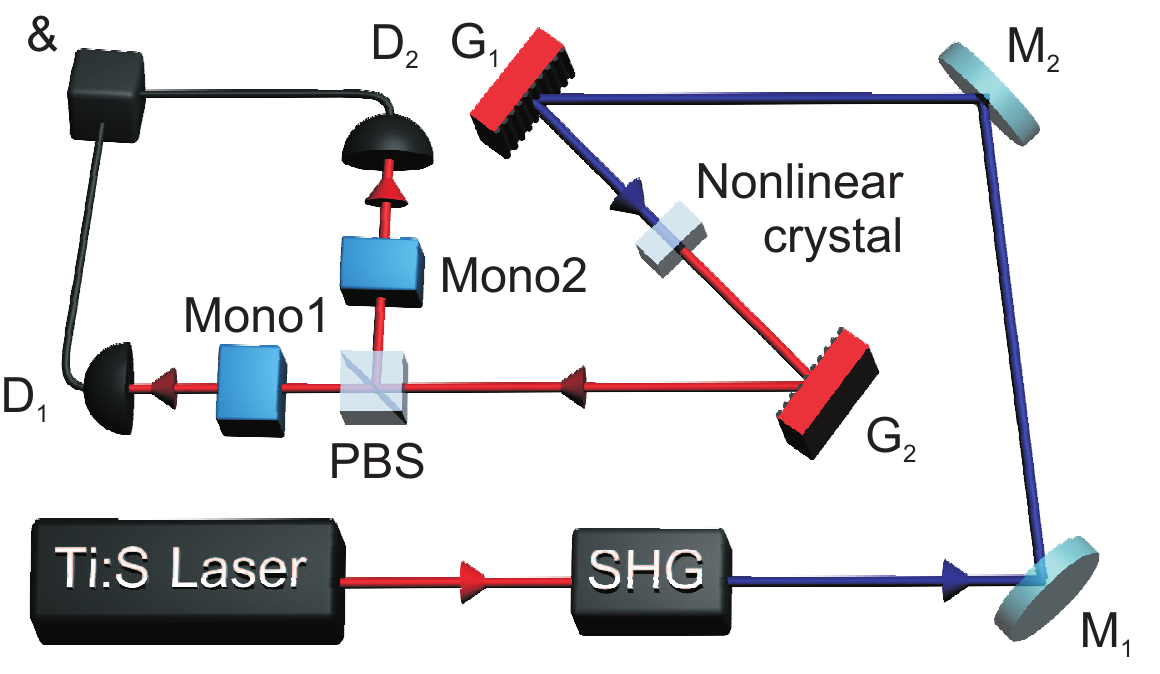}}}
\caption{(color online). Experimental setup. A $3.5$ mm BBO
crystal is pumped by the second harmonic of a femtosecond
Ti:sapphire laser. M: dichroic mirrors. G: diffraction gratings.
PBS: polarization beam splitter. Mono: monochromators. D:
single-photon counting modules. \&: coincidence electronics.}
\label{setup}
\end{figure}

To demonstrate the full tunability of the type of frequency
correlations by means of angular dispersion, an experiment was
performed. As depicted in Fig.~\ref{setup}, a $3.5$ mm BBO crystal
cut for degenerate collinear type-II phase-matching was placed
between a pair of diffraction gratings: G1 in the path of the pump
photons and G2 in the path of the downconverted photons. The first
grating introduces the pulse-front tilt and the second grating
removes the angular dispersion from the downconverted beam. The
crystal was pumped by the second harmonic (220 mW of average
power) of a femtosecond Ti:sapphire laser tuned at $800$ nm. The
measured bandwidth (FWHM) of the pump beam at 400 nm was $\Delta
\lambda_p=2$ nm. After the second grating, the downconverted
photons were separated by a polarizing beam splitter (PBS) and
collected into multimode fibers. To perform the direct measurement
of the joint spectral intensity $S \left(
\omega_s,\omega_i\right)$, the signal and idler photons were sent
to two monochromators (Jobin Yvon MicroHR) Mono1 and Mono2,
respectively. The outputs of the monochromators were coupled into
optical fibers and detected by single-photon counting modules
(Perkin-Elmer SPCM-AQR-14-FC) in order to record singles and
coincidence counts between the two detectors. The joint spectral
intensity is measured by scanning the central wavelengths of the
bandpass of the monochromators.

The effect of the gratings G1 and G2 is to introduce the
appropriate amount of angular dispersion, or equivalently
pulse-front tilt $\xi$, into the pulsed pump and the down
converted photons. The tilt is introduced in the plane determined
by the pump beam and the optic axis of the nonlinear crystal. The
gratings were chosen and placed in the setup in such a way that
they introduce opposite angular dispersion; this guaranties that
the tilt only modifies $\Delta_k$. The groove densities of G1 and
G2 were 1200 and 600 grooves per mm, resp., in the anticorrelated
and uncorrelated cases. In the correlated case, the same pair of
gratings could have been used, but in order to improve the
diffraction efficiency, a pair of gratings with 2400 and 1200
grooves per mm were used instead.

\begin{figure}[ht]
\centerline{\scalebox{1}{\includegraphics{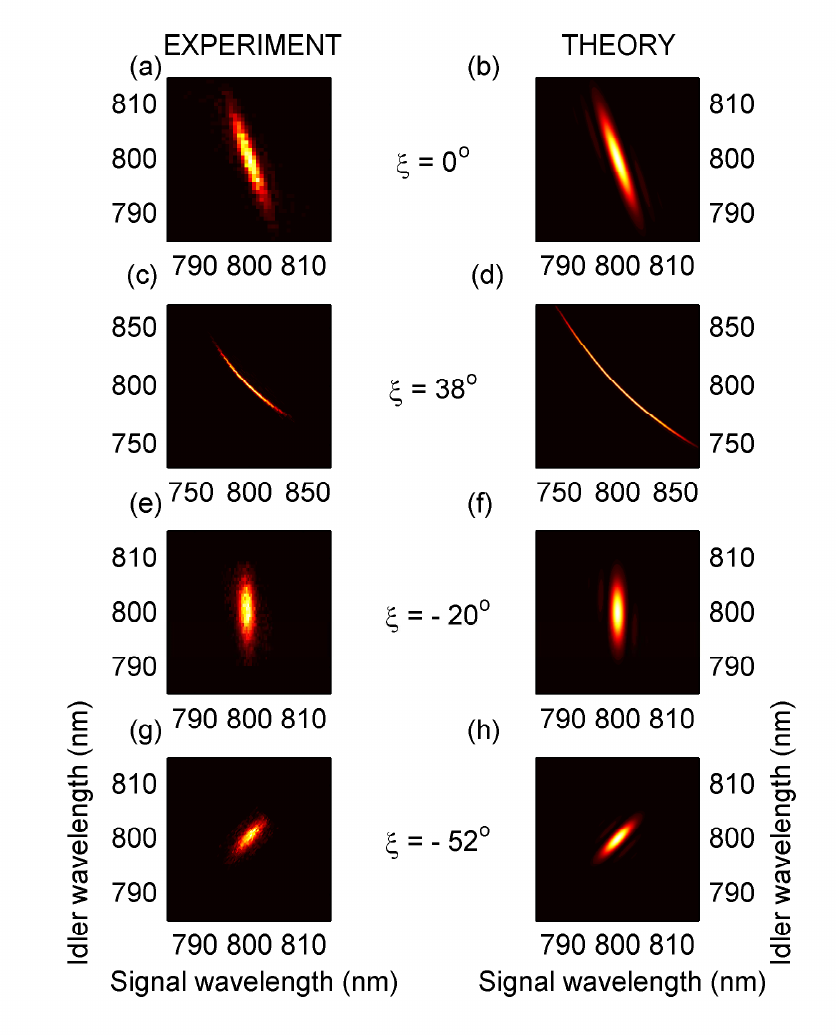}}}
\caption{(color online). Shape of the frequency correlations.
Experiment: left column; theoretical prediction: right column. (a)
and (b): no tilt, $\xi=0^{o}$; (c) and (d): anticorrelated
photons, $\xi=38^{o}$; (e) and (f): uncorrelated photons,
$\xi=-20^{o}$; (g) and (h): correlated photons
$\xi=-52^{o}$.}\label{experi}
\end{figure}

The left column of Fig.~\ref{experi} displays the experimental
measurements that demonstrate the full tunability of the frequency
correlations of paired photons. The different joint spectral
intensities were measured in the same experimental setup at the
same pump wavelength and with the same crystal. The only varying
parameter was the amount of pulse-front tilt $\xi$ introduced by
the gratings. Practically, the tilt tuning was achieved by
changing the angle of incidence at the gratings. For the sake of
comparison, the theoretical predictions are depicted in the right
column. They are obtained by plotting
$S\left(\omega_s,\omega_i\right) =
    |E_p \left(\omega_s+\omega_i\right)|^2
     \rm{sinc}^2 \left(\Delta_k L/2 \right)$,
where the pump frequency distribution is assumed to be Gaussian
with a FWHM bandwidth of $2$ nm and $\Delta_k$ is expanded up to
the second order in a Taylor series. The second order, however,
introduces only a slight correction as it is the first order that
dominates in type-II phase-matching.

The first row of Fig.~\ref{experi} shows the case with no tilt. As
expected for a pulsed pump and type-II phase-matching, the spectra
of the signal and idler are different, one being narrower than the
other. This is due to the fact that the downconverted photons have
different polarizations and by extension different group
velocities and group velocity dispersion. The joint spectrum lies
neither along the diagonal ($+ 45^o$), nor the antidiagonal line
($- 45^o$). The distinguishability between the spectra of the
signal and idler photons results in low visibility in interference
experiments with pulsed sources \cite{distinguishability}.
\\
\indent The second row of Fig.~\ref{experi} depicts
frequency-anticorrelated photons, the same type of frequency
correlations that would be obtained with a continuous-wave pump,
but in this case the entangled photon pairs were generated with a
broadband pump. The pulse-front tilt needed to produce these
extremely highly entangled and indistinguishable photons was
$\xi=38^{o}$. The possibility to generate frequency-anticorrelated
photons with pulsed lasers without the need for filtering is of
particular interest in order to obtain high-visibility
interference in experiments when the timing information provided
by the pump is crucial. This type of frequency correlations
exhibit high visibility in a Hong-Ou-Mandel interferometer
\cite{Martin_OL07}. The increase in the singles' bandwidth to
$\sim 90$ nm results in theoretical temporal correlations of the
entangled photons of $\sim 12$ fs \cite{hendrych_PRA09}.
Furthermore, the entropy of entanglement is boosted as well, as it
is a function of the ratio of the bandwidth of the downconverted
photons to the bandwidth of the pump \cite{plenio_PRA00,
hendrych_PRA09}.

The third row of Fig.~\ref{experi} depicts the case of
frequency-uncorrelated photons $S\left(\omega_s,\omega_i\right) =
S_s(\omega_s) S_i(\omega_i)$ that was obtained for $\xi = -20^o$.
We performed the Schmidt decomposition of the state given by Eq.
(\ref{jointspectrum}) that showed the entropy of entanglement to
be nearly 0 \cite{torres1}. Therefore, the frequency uncorrelation
observed in Fig.~\ref{experi}(e) is indeed a signature of the
presence of a separable quantum state.

To quantify the degree of frequency uncorrelation of our
two-photon state, we fit the experimental data with a Gaussian
function of the form
$\exp(-a\Omega_{s}^2-b\Omega_{i}^2-2c\Omega_{s}\Omega_{i})$
\cite{plenio_PRA00}. $c^2/(ab)$ that describes its orientation and
ellipticity, takes on values between 0 and 1 that correspond to
frequency-uncorrelated and maximally entangled states, resp. The
obtained value is $c^2/(ab) = 0.01$ with an overlap of the fit and
the experimental data of 98.1\%.

\begin{figure}[t]
\centerline{\scalebox{0.93}{\includegraphics{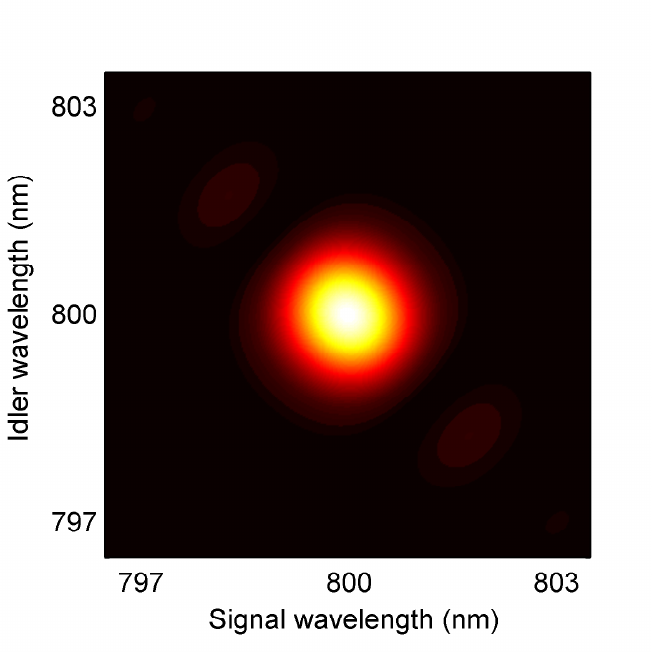}}}
\caption{(color online). Theoretical joint spectral intensity of
frequency-uncorrelated, indistinguishable photons.} \label{circle}
\end{figure}

Figures~\ref{experi}(e) and (f) show photons with distinguishable
spectra. It is worth mentioning, however, that a circular
distribution of the joint spectrum can also be obtained for a
type-II crystal with the help of angular dispersion (see
Fig.~\ref{circle}). If the bandwidth of the pump is reduced to 0.5
nm (e.g., by an interference filter), frequency-uncorrelated, and
in this case also indistinguishable, photons are generated. Their
bandwidth is $\Delta \lambda_{uncorr} = 2\sqrt(2)\Delta \lambda_p
= 1.4$ nm. The case where the two photons are frequency
uncorrelated and have the same bandwidth is of particular
interest, because it provides a source of heralded,
indistinguishable single photons for quantum communications and
processing applications \cite{singlephotons,mosley1}. Also, this
source would allow the implementation of superpositions of quantum
operators to generate different quantum states~\cite{Bellini}.

The last row of Fig.~\ref{experi} depicts the direct measurement
of photons with frequency correlation.  Unlike in
Ref.~\cite{kuzucu2}, where the nonlinear crystal and the
wavelength were chosen such as to generate frequency-correlated
photons, here we obtain the frequency correlation by applying a
pulse-front tilt of $\xi = -52^o$ with no need for changing the
material, neither the wavelength. Photons with this type of
correlations exhibit high visibility in interferometric
applications, can be used in quantum metrology to enhance the
precision of measurements, and can enhance the transmission of
polarization entanglement in optical fibers with polarization mode
dispersion \cite{pmd}.
\\
\indent In conclusion, we have demonstrated experimentally the
tunability of the frequency correlations of paired photons by
applying pulse-front tilt with no further changes of the SPDC
source. The transition from frequency-anticorrelated to correlated
photons is clearly seen from the experimental data. As cases of
particular interest, we have shown the first direct measurement of
the joint spectra in which frequency-correlated photons, and
anticorrelated photons pumped by an ultrafast pump were generated.
We have also demonstrated the generation of frequency-uncorrelated
photons. In the case of anticorrelated photons, the bandwidth, and
thus the entropy of entanglement, were increased resulting in
theoretically extremely short temporal correlations. The results
reported here are of valuable interest, as the used technique
works independently of the wavelength and the nonlinear crystal,
and therefore it can be implemented in materials and at
wavelengths where conventional solutions do not hold.
\\
\indent We would like to appreciate fruitful discussions with G.
Molina-Terriza. This work was supported by European Commission
(QAP, contract 015848) and by Government of Spain (Consolider
Ingenio CSD2006-00019, FIS2007-60179). M.M. thanks to project MSM
6198959213.


\begin{thebibliography}{13}

\bibitem{singlephotons} I.A. Walmsley and M.G. Raymer, Science {\bf 307}, 1733 (2005),
L.M. Duan, M. Lukin, J.I. Cirac, and P. Zoller, Nature
\textbf{414}, 413 (2001).

\bibitem{metrology} V. Giovannetti, S. Lloyd, and L. Maccone, Science
\textbf{306}, 1330 (2004). V. Giovannetti, L. Maccone, and S.
Lloyd, Nature {\bf 412}, 417 (2001), A. Valencia, G. Scarcelli,
and Y.H. Shih, Appl.\ Phys.\ Lett.\, \textbf{85}, 2655 (2004).

\bibitem{distinguishability} T.E. Keller, M.H. Rubin, Phys.\ Rev.\ A {\bf 56}, 1534 (1997), W.P.
Grice and I.A. Walmsley, {\em ibid.} {\bf 56}, 1627 (1997), G. Di
Giuseppe, L. Haiberger, F. De Martini, and A.V. Sergienko, {\em
ibid.} {\bf 56}, R21 (1997).

\bibitem{mosley1} P.J. Mosley {\em et al.}, Phys.\
Rev.\ Lett.\ \textbf{100}, 133601 (2008).

\bibitem{kuzucu1} O. Kuzucu {\em et al.},
Phys.\ Rev.\ Lett.\ {\bf 94},  083601 (2005).

\bibitem{Martin_OL07} M. Hendrych, M. Mi\v{c}uda, and J. P. Torres, Opt.\ Lett.\ {\bf 32}, 2339
(2007).

\bibitem{Ale_PRL07} A. Valencia {\em et al.},
Phys.\ Rev.\ Lett.\ {\bf 99}, 243601 (2007).

\bibitem{Walmsley_PRL09} O.Cohen {\em et al.}, Phys.\ Rev.\ Lett.\ {\bf 102}, 123603 (2009).

\bibitem{torres1} J.P. Torres, F. Maci\`{a}, S. Carrasco, and L. Torner, Opt.\ Lett.\ {\bf 30}, 314 (2005).

\bibitem{kuzucu2} O. Kuzucu, F.N.C. Wong, S. Kurimura, and S. Tovstonog,
Phys.\ Rev.\ Lett.\ {\bf 101}, 153602 (2008).

\bibitem{UltrafastBook} J.C. Diels and W. Rudolph, Ultrashort
laser pulse phenomena, Chap. 2.6, Academic Press, San Diego, 1995.

\bibitem{torres2} J.P. Torres, S. Carrasco, L. Torner, and E.W. VanStryland, Opt.\ Lett.\ {\bf 25}, 1735 (2000).

\bibitem{hendrych_PRA09} M. Hendrych, X. Shi, A. Valencia, and J. P. Torres, Phys.\ Rev.\ A {\bf 79}, 023817 (2009).

\bibitem{plenio_PRA00} S. Parker, S. Bose, and M.B. Plenio, Phys.\ Rev.\ A {\bf 61}, 032305 (2000).

\bibitem{Bellini} V. Parigi, A. Zavatta, M. Kim, and M. Bellini, Science {\bf 317}, 1890 (2007).

\bibitem{pmd}P.S.Y. Poon and C.K. Law, Phys.\ Rev.\ A {\bf 77}, 032330 (2008).

\end{thebibliography}
\end{document}